\documentstyle[floats,prd,aps,epsfig,eqsecnum,12pt]{revtex}

\makeatletter
\newbox\tempboxa
\newdimen\captionboxsubcount
\def\capsize#1{\captionboxsubcount=#1pt}
\newdimen\captionboxsub
\captionboxsub=\hsize \advance\captionboxsub by -\captionboxsubcount
\advance\captionboxsub by -\captionboxsubcount
\long
\def\@makecaption#1#2{
 \setbox\@tempboxa\hbox{#1 #2}
 \ifdim \wd\@tempboxa >\captionboxsub
\rightskip=\captionboxsubcount \leftskip=\captionboxsubcount #1 #2
\else \hbox to\hsize{\hfil\box\@tempboxa\hfil}
 \fi}
\makeatother
\capsize{30}

\begin{document}

\vfill
\begin{titlepage}
\begin{flushright}
\begin{minipage}{5cm}
\begin{flushleft}
\small
\baselineskip = 10pt NORDITA-2000-114-HE \\ YCTP-P-10-00
\end{flushleft}
\end{minipage}
\end{flushright}

\begin{center}
\Large\bf \centerline{Electroweak Physics for Color
Superconductivity}
\end{center}
\footnotesep = 12pt
\vfill
\begin{center}
\large \centerline{Roberto {\sc Casalbuoni}$^{a}$
\footnote{Electronic address: {\tt Roberto.Casalbuoni@fi.infn.it}}
\quad Zhiyong {\sc Duan}$^b$ \footnote{Electronic address: {\tt
zhiyong.duan@yale.edu}} \quad Francesco {\sc Sannino}$^c$
\footnote{ Electronic address: {\tt francesco.sannino@nbi.dk}}}
\vskip .5cm $^{a}${\it  Dipartimento Di Fisica, Univ. di Firenze
and I.N.F.N Sezione di Firenze,  I-50125, Italia.} \vskip .5cm
$^b${\it Department of Physics, Yale Univ., New Haven,  CT
06520-8120, USA.} \vskip .5cm $^c${\it NORDITA, Blegdamsvej 17
DK-2100 Copenhagen \O, Denmark.}
\end{center}
\vfill
\begin{center}
\bf Abstract
\end{center}
\begin{abstract}
\baselineskip = 17pt We construct the effective theories
describing the electroweak interactions for the low energy
excitations associated with the color superconductive phases of
QCD at high matter density. The main result, for the 3 flavor
case, is that the quasiparticle Goldstone boson $\pi^0$ decay into
two physical massless photons is identical to the zero density
case once we use the new Goldstone decay constant and the modified
electric charge $\widetilde{e}=e\, \cos\theta$, with $\tan\theta
=2e/\sqrt{3}g_s$ and $g_s$ the strong coupling constant. {}For 2
flavors we find that the coupling of the quarks to the neutral
vector boson $Z^0$ is modified with respect to the zero density
case. We finally point out possible applications of our result to
the physics of compact objects.
\end{abstract}
\begin{flushleft}
\footnotesize
\end{flushleft}
\vfill
\end{titlepage}

\section{Introduction}

\label{uno}

Recently quark matter at very high density has attracted a great
flurry of interest
\cite{ARW_98,RSSV_98,ARW_98b,charges,REV,SW,SW_98b}. In this
limit, quark matter is expected to behave as a color
superconductor \cite{ARW_98,RSSV_98}. Possible phenomenological
applications are associated with the description of neutron star
interiors, neutron star collisions and the physics near the core
of collapsing stars.

In a superconductive phase, the color symmetry is spontaneously
broken and a hierarchy of scales, for given chemical potential, is
generated. Indicating with $g_s$, the underlying coupling
constant, the relevant scales are: the chemical potential $\mu $
itself, the dynamically generated gluon mass $m_{gluon}\sim g_s\mu
$ and the Gap parameter $\Delta \sim \frac{\mu }{g_s^{5} }
e^{-\frac{\alpha }{g_s}}$ with $\alpha$ a calculable constant.
Since for high $\mu$ the coupling constant $g_s$ (evaluated at the
fixed scale $\mu $) is $\ll 1$ , we have:
\begin{equation}
\Delta \ll m_{gluon}\ll \mu \ .
\end{equation}
Massless excitations dominate physical processes at very low
energy with respect to the energy Gap ($\Delta$). Their spectrum
is intimately related to the underlying global symmetries and the
way they are realized at low energies. They also obey low energy
theorems governing their interactions which can be usefully
encoded in effective low energy Lagrangians (like for cold and
dilute QCD \cite{ChPT}). It is possible to order the effective
Lagrangian terms describing the Golstone boson self interactions
in number of derivatives. The resulting theory for dilute QCD is
named Chiral Perturbation Theory \cite{ChPT}.  Unfortunately this
well defined scheme is not sufficient for a complete description
of hadron dynamics since new massive hadronic resonances appear at
relatively low energies like the $\sigma$ or the vector $\rho$ and
new effective Lagrangians of the type described in \cite{effective
Lagrangians} are needed.

 Another set of relevant constraints is provided by 't Hooft
anomaly matching conditions \cite{tHooft}. In Reference \cite{S},
it was shown that the low energy spectrum, at finite density,
displays the correct quantum numbers to saturate the 't~Hooft
global anomalies. It was also observed that QCD at finite density
can be envisioned, from a global symmetry and anomaly point of
view, as a chiral gauge theory \cite{ball,ADS}. In Reference
\cite{HSaS} it was demonstrated, using a variety of field
theoretical tools, that 't Hooft anomaly matching conditions must
hold for any cold but dense gauge theory.

In this paper we construct the electroweak interactions to
complete the low energy effective theory describing Quantum Chromo
Dynamics with two and three flavors at high density. This work can
be considered as the first step for properly describing the
phenomenology associated with electroweak processes stemming, for
example, from the core of some neutron stars which may be dense
enough to be in a two, three or both flavor superconductive phases
\cite{NS}.

Some possible implications of electroweak
interactions in compact objects for 3 flavors QCD at high matter density
has been also partially
investigated in \cite{HLNR}.

The low energy theory for two flavors without considering
electroweak interactions is provided in Ref.~\cite{CDS} while the
one for three flavors has been developed in \cite{CG}. We use the
non linear framework \cite{CWZ} to construct the low energy
theories.

First for the three flavor case, we include the $QED$ interactions
and complete the low energy theory by implementing the global
anomalies via the Wess-Zumino-Witten term. This topological term
is essential when describing the time honored $\pi ^{0}\rightarrow
\gamma \gamma $ process at finite density \cite{HRZ}.

By considering in detail the axial anomaly at the fundamental
level and by comparing it with the gauged version of the
Wess-Zumino term we demonstrate that the coefficient of the
topological term is the same as zero density. The only place where
the finite density effects enters are in the redefinition of the
electric charge which is now given by $e\cos\theta$ with
$\tan\theta=2e/\sqrt{3}g_s$ and in the high density value of the
Goldstone bosons decay constant. Our result stems directly from
the Higgsing nature of the Color superconductive phase and it is
at variance with the result presented in \cite{HRZ}. We briefly
comment on the nature of the discrepancy.

Besides the importance of the $\pi^0\rightarrow \gamma \gamma$
process per se and its phenomenological consequences  when
investigating possible Quark-type stars featuring a CFL core, the
topological coefficient term is also critical when considering the
solitonic (Skyrme) solutions of the effective Lagrangian. In fact
now we have the same winding number as for ordinary QCD and hence
we get massive excitations, which after collective quantization,
describe spin half particles with quantum numbers identical to the
ordinary baryons. This is the actual realization of the
Quark-Hadron continuity \cite{SW} advocated for the baryon-type
sector of the Color-superconductive phase. We then introduce the
electroweak interactions for 3 flavors.

{} Next we generalize the effective Lagrangian theory for 2
flavors presented in \cite{CDS} to describe electroweak
interactions. We observe that the Lagrangian not only reproduces
the physical eigenstates for the photon and the eighth gluon
\cite{charges} but also predicts a modified coupling of the quarks
to the neutral vector boson $Z^0$ with respect to the zero density
case. The physical effect, which is our main result for the 2
flavor case, is not suppressed with respect to the zero density
weak processes. On general grounds we expect this result to affect
the Quark-stars (with a 2SC component) cooling processes
\cite{Shapiro} via neutrino emission. In particular we have in
mind the neutrino transparency question directly related to the
mean free path of a neutrino in a compact star. This question has
already attracted some interest \cite{CR}. Indeed for ordinary
neutron stars \cite{Shapiro} the mean free path is significantly
affected by the $neutron-{\nu}$ scattering reaction due to the
presence of neutral currents. In a 2SC star we then expect the
$quark-{\nu}$ scattering to play an equally important role.

In Sect. \ref{3f} we study the anomalous process
$\pi^0\rightarrow\gamma \gamma$ at high matter density, complete
the effective Lagrangian for the quasiparticle Goldstone bosons by
adding the Wess-Zumino term and comment on related issues. In
Sect. \ref{weak3} we gauge the weak sector for the three flavor
case. In Sect. \ref{tre} we briefly review the 2 flavor case while
the generalization to include the electroweak processes is done in
Sect. \ref{quattro}. Finally in Sect. \ref{cinque} we summarize
and conclude.

\section{QED for the 3 flavor Case: the $\pi_0\rightarrow \gamma \gamma$ process}

\label{3f}

Let us start with the case of $N_{f}=3$ light flavors. At zero density only
the confined Goldstone phase is allowed and the resulting symmetry group is
$SU_{V}(3)\times U_{V}(1)$. Indeed there is no solution for the 't
Hooft anomaly conditions with massless composite fermions leaving
intact the flavor group. In this case the topological Wess-Zumino
term for the Goldstone bosons is needed to implement the global
anomalies of the underlying theory at the effective Lagrangian
level. The Vafa-Witten theorem \cite{VW}, valid for vector-like
theories, prohibits the further breaking of the remaining
vector-like symmetries like $U_{V}(1)$.

Turning on low baryon density we expect the theory to remain in the confined
phase with the same number of Goldstone bosons (i.e. 8). Evidently the 't
Hooft anomaly conditions are still satisfied. At very high density,
dynamical computations suggest \cite{ARW_98b} that the preferred phase is a
superconductive one and the following ansatz for a quark-quark type of
condensate is energetically favored:
\begin{equation}
\epsilon ^{\alpha \beta }<q_{L\alpha ;a,i}q_{L\beta ;b,j}>\sim k_{1}\delta
_{ai}\delta _{bj}+k_{2}\delta _{aj}\delta _{bi}\ ,  \label{condensate}
\end{equation}
\noindent and a similar expression holds for the right transforming fields.
The Greek indices represent spin, $a$ and $b$ denote color while $i$ and $j$
indicate flavor. The condensate breaks the gauge group completely while
locking the left/right transformations with color. The final global symmetry
group is $SU_{c+L+R}(3)$, and the low energy spectrum consists of $9$
Goldstone bosons. Before constructing the Lagrangian for the true massless
Goldstones it is instructive to introduce the left and right transforming
fields \cite{CG}:
\begin{equation}
L\rightarrow g_{L}Lg_{c}\ ,\qquad R\rightarrow g_{R}Rg_{c}\ ,
\end{equation}
where $L/R$ parameterizes the Goldstone bosons induced by the appearance of
the condensate (\ref{condensate}). $g_{L/R}\in SU_{L/R}(3)$ while $g_{c}\in
SU(3)$ of color. The covariant derivative describing color and
electromagnetic interactions is
\begin{equation}
D_{\mu }L=\partial _{\mu }L-i\,eA_{\mu }\,QL-ig_{s}\,G_{\mu }^{m}LT^{m}\ ,
\label{cov1}
\end{equation}
\-with $Q={\rm {diag}\left( 2/3,-1/3,-1/3\right) }$ the ordinary
quark charges, and $T^{m}$ the generators for color which, for
convenience, are defined such that $T^{8}=\frac{\sqrt{3}}{2}Q$ and
$T^{3}={\rm {diag}(0,\frac{ 1}{2},-\frac{1}{2})}$. Likewise for
the $R$ field. Here $A$ denotes the standard photon field while
$G^{m}$ are the gluon fields. The vev in Eq.~(\ref{condensate})
(corresponding to $L\propto \delta _{ic}$) locks together flavor
and color and the Higgs mechanism sets in providing masses for all
but one linear combination of the gluon $G^{8}$ and photon $A$.
The massive eigenstate can be easily identified by investigating
the kinetic term
\begin{equation}
{\rm Tr}\left[ D_{\mu }L^{\dagger }D^{\mu }L\right]  \label{kinetic}
\end{equation}
which leads to a mass term of the type:
\begin{equation}
g_{s}^{2}\frac{1}{2}\sum_{m=1}^{7}{G^{m}}^{2}+{\rm Tr}\left[ Q^{2}\right]
\left( e\,A+\frac{\sqrt{3}}{2}g_{s}G^{8}\right) ^{2}\ ,
\end{equation}
where for simplicity we have suppressed the Lorentz indices. We can now
identify the massive $\widetilde{G}^{8}$ and the orthogonal massless
eigenstate $\widetilde{A}$ via:
\begin{eqnarray}
\widetilde{G}^{8} &=&\cos \theta G^{8}+\sin \theta A\ , \\
\widetilde{A} &=&-\sin \theta G^{8}+\cos \theta A\ ,
\end{eqnarray}
with
\begin{equation}
\cos \theta =\sqrt{3}\frac{g_{s}}{\sqrt{3g_{s}^{2}+4e^{2}}}\ ,\qquad \sin
\theta =2\frac{e}{\sqrt{3g_{s}^{2}+4e^{2}}}\ .
\end{equation}
$\widetilde{A}$ is reinterpreted as the physical photon.

The same covariant derivative for the underlying quark fields can be
compactly written as:
\begin{equation}
D_{\mu }=\partial _{\mu }-ie\,A_{\mu }Q\times {\mbox{\bf
1}}-ig_{s}\,G_{\mu }^{m}{\mbox {\bf 1}}\times T^{m}\ .
\end{equation}
The first $3\times 3$ matrix is in flavor space while the second is in
color. This notation is also convenient when investigating the global
anomalies at the fundamental fermion level. The unbroken generator
associated with the new photon $\widetilde{A}$ is, in flavor$\times $ color
space,
\begin{equation}
\widetilde{Q}=Q\times {\mbox {\bf 1}}-{\mbox {\bf 1}}\times Q\ .
\end{equation}
It is easy to check that all quarks, in medium, have integer
charges in $\widetilde{e}$ units\footnote{Clearly with respect to
the new photon field $\widetilde{A}$ the new charge units is $e
\cos \theta$ and the ratio between the quark and electron charge
is integer.} and the condensate (\ref{condensate}) is indeed
neutral under $\widetilde{Q}$.

 The
relevant terms in the covariant derivative can be rewritten as:
\begin{equation}
g_{s}{G}^{8}\,{\bf 1}\times {T}^{8}+eA\,Q\times {\bf 1}=g_{s}^{\prime }
\widetilde{G}^{8}\left[ \frac{\sqrt{2}}{2}\widetilde{T}^{8}-\frac{\sqrt{3}}{
4 }\widetilde{Q}\cos (2\theta )\right] +\widetilde{e}\widetilde{A}\widetilde{
Q} \ ,
\end{equation}
where $\widetilde{e}=e\cos \theta $ and $g_{s}^{\prime }=g_{s}/\cos \theta$.
The second term corresponds to the unbroken $U(1)$ gauge symmetry which we
identify as the new QED with coupling $\widetilde{e}$. $\widetilde{T}^{8}$
is the generator orthogonal to $\widetilde{Q}$, normalized according to
${\rm Tr}\left[ \widetilde{T}^{8}\widetilde{T}^{8}\right] =3/2$ and is given
by:
\begin{equation}
\widetilde{T}^{8} =\frac{\sqrt{6}}{4}(Q\times {\mbox {\bf 1}}+\frac{2\sqrt{3}}{3}{\mbox {\bf 1}}\times {T}^{8}) \equiv\frac{\sqrt{6}}{4}(Q\times {\mbox
{\bf 1}}+{\mbox {\bf 1}}\times Q\ )\ .
\end{equation}

 It is well known that anomalies play an important role for
describing phenomenological processes like $\pi _{0}\rightarrow
\gamma \gamma $ in cold and dilute matter. But before discussing
the anomaly equation we should understand how the physical
Goldstone bosons emerge. At high density, in the 3 flavor case,
the symmetry group $SU_{L}(3)\times SU_{R}(3)\times SU_{c}(3)$
breaks spontaneously to $SU_{c+L+R}(3)$ leaving behind $16$
Goldston bosons. However, being $SU_c(3)$ a gauge group 8
Goldstone bosons are absorbed in the longitudinal components of
the massive gluons. So we are left with 8, not colored, physical
massless Goldstone bosons. They can be encoded in the unitary
matrix \cite{CG}
\begin{equation}
U=LR^{\dagger } \ ,
\end{equation}
transforming linearly under the left-right flavor rotations
\begin{equation}
U\rightarrow g_{L}Ug_{R}^{\dagger }\ .
\end{equation}
with $g_{L/R}\in SU_{L/R}(N_{f})$. In our notation $U$ is the transpose of $
\Sigma $ defined in Ref.~\cite{CG}. $U$ satisfies the non linear realization
constraint $UU^{\dagger }=1$. We also require ${\rm det}U=1$. In this way we
avoid discussing the axial $U_{A}(1)$ anomaly at the effective Lagrangian
level. (See Ref.~\cite{SS} for a general discussion of trace and $U_{A}(1)$
anomaly). We have
\begin{equation}
U=e^{i\frac{\Phi }{F}}\ ,
\end{equation}
with $\Phi =\sqrt{2}\Phi ^{a}t^{a}$ representing the $8$ Goldstone bosons. $
t^{a}$ are the standard generators of $SU(3)$ (i.e. $t^{3}={\rm {diag}(\frac{
1}{2},-\frac{1}{2},0)}$), with $a=1,...,8$ and $\displaystyle{{\rm Tr}\left[
t^{a}t^{b}\right] =\frac{1}{2}\delta ^{ab}}$. $F$ is the Goldstone bosons
decay constant at finite density. Clearly $\pi^0$ is associated with $t^3$.

Recently it has been shown in \cite{S} that all of the interesting
superconductive phases do respect, as for general chiral gauge theories at
zero density, global anomaly matching conditions a la 't Hooft. In \cite
{HSaS} it has been shown that anomaly matching conditions are unmodified
with respect to the zero density case provided that one correctly implements
the group theoretical structure of the global anomalies. At high density, in
the 3 flavor case, using the fact \cite{S,HSaS} that the anomalous
coefficient is unmodified by density effects, we have that the anomalous
variation of the axial current $j_{5}^{{\mu }3}$ associated with $\pi ^{0}$
is given by:
\begin{eqnarray}
\partial _{\mu }j_{5}^{{\mu }3} &=&-\frac{e^{2}}{16\pi ^{2}}\epsilon
^{\alpha \beta \mu \nu }F_{\alpha \beta }F_{\mu \nu }{\rm Tr}[(t^{3}\times
{\bf 1)}(Q\times {\bf 1})^{2}]  \nonumber \\
~ &=&-\frac{e^{2}N_{c}}{16\pi ^{2}}\epsilon ^{\alpha \beta \mu \nu
}F_{\alpha \beta }F_{\mu \nu }{\rm Tr}[t^{3}Q^{2}]  \nonumber \\
~ &=&-\frac{e^{2}}{32\pi ^{2}}\epsilon ^{\alpha \beta \mu \nu }F_{\alpha
\beta }F_{\mu \nu }\ ,
\end{eqnarray}
where $N_{c}=3$ comes from the trace over color space and $F_{\alpha \beta
}=\partial _{\alpha }A_{\beta }-\partial _{\beta }A_{\alpha }=\cos \theta \;\widetilde{F}_{\alpha \beta }+\sin \theta \;\widetilde{G}_{\alpha \beta
}^{8} $ with $\widetilde{F}_{\alpha \beta }=\partial _{\alpha }\widetilde{A}_{\beta }-\partial _{\beta }\widetilde{A}_{\alpha }$ and $\widetilde{G}_{\alpha \beta }^{8}=\partial _{\alpha }\widetilde{G}_{\beta }^{8}-\partial
_{\beta }\widetilde{G}_{\alpha }^{8}$. This leads to the following
expression in terms of the physical vector eigenstates:
\begin{eqnarray}
\partial _{\mu }j_{5}^{{\mu }3} &=&-\frac{e^{2}}{32\pi ^{2}}\epsilon
^{\alpha \beta \mu \nu }F_{\alpha \beta }F_{\mu \nu }  \nonumber \\
&=&-\frac{e^{2}}{32\pi ^{2}}\epsilon ^{\alpha \beta \mu \nu }(\cos
^{2}\theta \widetilde{F}_{\alpha \beta }\widetilde{F}_{\mu \nu }+2\sin
\theta \cos \theta \widetilde{F}_{\alpha \beta }\widetilde{G}_{\mu \nu
}^{8}+\sin ^{2}\theta \widetilde{G}_{\alpha \beta }^{8}\widetilde{G}_{\mu
\nu }^{8})\ .  \label{anomaly}
\end{eqnarray}

\begin{figure}[th]
\begin{center}
\input{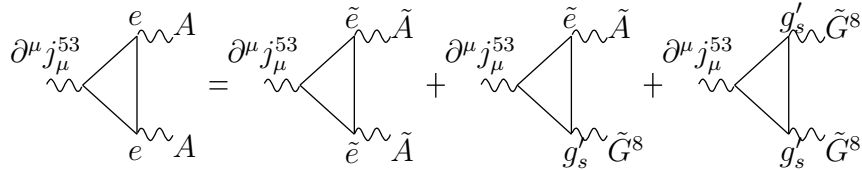}
\end{center}
\caption{Triangle anomaly for 3 flavors QCD at high density.}
\label{feynman}
\end{figure}

In figure \ref{feynman} we graphically represent equation (\ref{anomaly})
via Feynman diagrams. We have ${g_{s}^{\prime }=\frac{2\sqrt{3}}{3}\frac{e}{\sin \theta }}$ and $\widetilde{e}=e\cos \theta $. {}For each $\widetilde{A}$
we associate the generator $\widetilde{Q}=Q\times {\bf 1}-\frac{2\sqrt{3}}{3}
{\bf 1}\times {T} ^{8}$ and for $\widetilde{G}^{8}$ the generator $\frac{
\sqrt{2}}{2} \widetilde{T}^{8}-\frac{\sqrt{3}}{4}\widetilde{Q}\cos (2\theta
)=\cos ^{2}\theta \,{\bf 1}\times {T}^{8}+\frac{\sqrt{3}}{2}\sin ^{2}\theta
\,Q\times {\bf 1}$ as prescribed by the covariant derivative. Clearly the
first diagram, in the right hand side, corresponds to $\pi ^{0}\rightarrow
\widetilde{\gamma }\widetilde{\gamma }$ in the superconducting phase where $\widetilde{\gamma }$ indicates the physical massless photon $\widetilde{A}$.
We conclude that the anomalous electromagnetic properties of the
superconductive state are identical to the ones of ordinary $QCD$ provided
that we replace the electric charge $e$ with $\widetilde{e}=e\cos \theta$.

We are now ready to review the low energy effective theory for the 3 flavor
case at high matter density. The effective Lagrangian at low energies \cite
{CG} for the true massless Goldstone bosons is similar to the ordinary
effective Lagrangian for QCD at zero density except for an extra Goldstone
boson associated with the spontaneously broken $U_{V}(1)$ symmetry which we
will not consider for the moment.

The effective Lagrangian globally invariant under chiral rotations is (up to
two derivatives and counting $U$ as a dimensionless field)
\begin{equation}
L=\frac{F^{2}}{2}\,{\rm Tr}\left[ \partial _{\mu }U\partial ^{\mu
}U^{\dagger }\right] \ .  \label{ef}
\end{equation}
The Wess-Zumino term \cite{WZ} can be compactly written by using
the language of differential forms. It is useful to introduce the
algebra valued Maurer-Cartan one forms:
\begin{equation}
\alpha =\left( \partial _{\mu }U\right) U^{-1}\,dx^{\mu }\equiv
\left( dU\right) U^{-1}\ , \qquad \quad
\beta=U^{-1}dU=U^{-1}\alpha U \ ,  \label{MC}
\end{equation}
which transform, respectively, under the left and right $SU(N_f)$
flavor group.
 The
Wess-Zumino effective action is
\begin{equation}
\Gamma _{WZ}\left[ U\right] =C\,\int_{M^{5}}{\rm Tr}\left[ \alpha ^{5}\right]
\ .  \label{WZ}
\end{equation}
The price to pay in order to make the action local is to augment
by one the space dimensions. Hence the integral must be performed
over a five-dimensional manifold whose boundary ($M^{4}$) is the
ordinary Minkowski space. We now show that the constant $C$ is the
same as the one at zero density, i.e.:
\begin{equation}
C=-i\frac{N_{c}}{240\pi ^{2}}\ , \label{coef}
\end{equation}
where $N_{c}$ is the number of colors (fixed to be 3 in this
case). This is a consequence of Eq.~(\ref{anomaly}). More
specifically we now compare the current algebra prediction for the
time honored process $\pi ^{0}\rightarrow 2\widetilde{\gamma }$
with the amplitude predicted using Eq.~(\ref{WZ}) by  gauging the
electromagnetic sector \cite{Witten,anomalies} of the Wess-Zumino
term.

Before gauging the Wess-Zumino term we need to stress that in
presence of an Higgsing phenomenon gauge symmetry is clearly not
lost and one is {\it always} entitled to work with the un-rotated
gauge fields and to rotate them to the mass engenstates only in
the very end.

In order to better understand the anomalies we recall that the
fully gauged Wess Zumino term under the $SU_L(N_f)\times
SU_R(N_f)$ chiral symmetry group is:
\begin{eqnarray}
\Gamma _{WZ}\left[ U,\;A_{L},\;A_{R}\right] &=&\Gamma _{WZ}\left[
U\right] \,+\,5Ci\,\int_{M^{4}}{\rm Tr}\left[ A_{L}\alpha
^{3}+A_{R} \beta ^{3}\right] \nonumber \\ &&-5C\,\int_{M^{4}}{\rm
Tr}\left[ (dA_{L}A_{L}+A_{L}dA_{L})\alpha
+(dA_{R}A_{R}+A_{R}dA_{R})\beta \right]  \nonumber \\
&&+5C\,\int_{M^{4}}{\rm Tr}\left[
dA_{L}dUA_{R}U^{-1}-dA_{R}dU^{-1}A_{L}U \right]  \nonumber \\
&&+5C\,\int_{M^{4}}{\rm Tr}\left[ A_{R}U^{-1}A_{L}U\beta
^{2}-A_{L}UA_{R}U^{-1}\alpha ^{2}\right]  \nonumber \\
&&+\frac{5C}{2}\,\int_{M^{4}}{\rm Tr}\left[ (A_{L}\alpha
)^{2}-(A_{R}\beta )^{2}\right] +5Ci\,\int_{M^{4}}{\rm Tr}\left[
A_{L}^{3}\alpha +A_{R}^{3}\beta \right]  \nonumber \\
&&+5Ci\,\int_{M^{4}}{\rm Tr}\left[
(dA_{R}A_{R}+A_{R}dA_{R})U^{-1}A_{L}U-(dA_{L}A_{L}
+A_{L}dA_{L})UA_{R}U^{-1} \right]  \nonumber \\
&&+5Ci\,\int_{M^{4}}{\rm Tr}\left[ A_{L}UA_{R}U^{-1}A_{L}\alpha
+A_{R}U^{-1}A_{L}UA_{R}\beta \right]  \nonumber \\
&&+5C\,\int_{M^{4}}{\rm Tr}\left[
A_{R}^{3}U^{-1}A_{L}U-A_{L}^{3}UA_{R}U^{-1}+\frac{1}{2}
(UA_{R}U^{-1}A_{L})^{2}\right]  \nonumber \\
&&-5Cr\,\int_{M^{4}}{\rm Tr}\left[ F_{L}UF_{R}U^{-1}\right] \ ,
\label{GWZ1}
\end{eqnarray}
with the two-forms $F_{L}$ and $F_{R}$ defined as
$F_{L}=dA_{L}-iA_{L}^{2}$ and $F_{R}=dA_{R}-iA_{R}^{2}$ and the
one form $A_{L/R}=A^{\mu}_{L/R}dx_{\mu}$. $r$ is a real arbitrary
parameter. The previous Lagrangian, when identifying the vector
fields with true gauge vectors, correctly saturates the underlying
global anomalies \cite{anomalies}.

The last term in Eq.~(\ref{GWZ1}) is a gauge covariant term.
However it is not invariant under parity\footnote{ {}For reader's
convenience we provide the rules for parity transformation
\begin{eqnarray}
&&A_{L,R}(\vec{x})\leftrightarrow A_{R,L}(-\vec{x})\;,\qquad
U(\vec{x} )\leftrightarrow U^{-1}(-\vec{x})\;,  \nonumber \\
&&\alpha \leftrightarrow -\beta \;,\qquad d\leftrightarrow
d\;,\qquad \mbox{(measure)}\leftrightarrow -\mbox{(measure)}\ ,
\nonumber 
\end{eqnarray}
as well as for charge conjugation
\begin{eqnarray}
A_{L,R}&\leftrightarrow& -A_{R,L}^{T}\;,\quad U\leftrightarrow
U^{T}\;,\quad \alpha \leftrightarrow \beta ^{T}\;.
\nonumber 
\end{eqnarray}} and so the parameter $r$ must vanish. All the other terms are
related by gauge invariance.

We can now restrict the attention to the electromagnetic
interactions by constraining the vector fields to satisfy the
equation
\begin{equation}
A_L=A_R=Q\, A=e\,Q\, A_{\mu} dx^{\mu} \ ,
\end{equation}
with $A_{\mu}$ the ordinary photon field. The Wess-Zumino term
collapses to the following \cite{anomalies} form:
\begin{eqnarray}
\Gamma _{WZ}\left[ U,\;A\right] &=&\Gamma _{WZ}\left[ U\right]
\,+\,5\, e\, Ci\,\int_{M^{4}} A\,{\rm Tr}\left[ Q\left(\alpha
^{3}+ \beta ^{3}\right)\right] \nonumber \\ &&-10\, e^2\,
C\,\int_{M^{4}}AdA{\rm Tr}\left[ Q^2\left(\alpha +\beta\right) +
\frac{1}{2}\left(QU^{-1}Q dU - QUQ dU^{-1}\right) \right] \ .
\label{WZelectro}
\end{eqnarray}
The last term leads to the following $\pi_0 \rightarrow \gamma
\gamma$ Lagrangian:
\begin{equation}
{\cal L}_{\pi_0\rightarrow\gamma\gamma}=-\frac{30}{4} e^2\, C\,i
{\rm Tr}\left[t^3 Q^2\right] \pi^0 \frac{\sqrt{2}}{F}
\epsilon^{\mu\nu\rho\sigma}F_{\mu\nu}F_{\rho\sigma} \ ,
\end{equation}
and $\pi^0=\Phi^3$. The last ingredient is the axial current
(expanded up to the first derivative) expressed as function of the
Goldstone bosons:
\begin{equation}
j_5^{\mu\,a}=-\frac{F}{\sqrt{2}}\partial^{\mu} \Phi^a \ .
\end{equation}
One pole saturation of the 3 point function anomalous amplitude
\cite{Peskinbook}, when compared with the underlying anomalous
variation in Eq.~(\ref{anomaly}) leads straightforwardly to the
result for $C$ in Eq.~(\ref{coef}).

One can also fix the Wess-Zumino coefficient by matching at the
effective Lagrangian level the anomalous variation for the left or
right currents associated with the non abelian {\it consistent}
\footnote{We differentiate here consistent anomaly from the
covariant anomaly\cite{RB}. The two anomalies are related. In
order to compare directly the axial anomalous variation without
using pole saturation it is convenient to consider the anomalies
in Bardeen's form. This is achieved by subtracting to the Wess
Zumino displayed in (\ref{GWZ1}) the action
$\Gamma_c=\Gamma_{WZ}\left[U=1;A_L,A_R\right]$ \cite{anomalies}.
The new action term (which clearly does not change the
$\pi^0\rightarrow \gamma \gamma$ process) reads
\begin{eqnarray}
\Gamma_{WZ}^{\prime}\left[U;A_L,A_R\right]&=&\Gamma_{WZ}\left[U;A_L,A_R\right]
-\Gamma_{WZ}\left[1;A_L,A_R\right]\ . \nonumber
\end{eqnarray}
If we restrict to the electromagnetic variations one gets
\cite{anomalies}
\begin{eqnarray}
\delta\left(\Gamma^{\prime}_{WZ}\right)&=&\frac{30}{2}\,e^2\, C \,
i \int_{M^4}\epsilon_5^3 {\rm Tr}\left[t^3 Q^2\right]\epsilon^{\mu
\nu \rho \sigma}F_{\mu\nu}F_{\rho\sigma} \
=-\int_{M^4}\epsilon_5^3
\partial_{\mu} j^{\mu\,3}_5 \nonumber \ ,
\end{eqnarray} which leads again to the result for $C$ obtained in the main text.} anomalies.

This procedure seems to be the one adopted in Ref.~\cite{HRZ}.
However in Ref.~\cite{HRZ} the anomaly coefficient (which should
be the same for all of the currents, left and right) differs from
our canonical one by a factor 3. The reason is that the physical
pions (i.e. quasiparticle) are colorless and hence a factor 3 (due
to the color factor) should appear in the anomalous coefficient.

More specifically, it seems to us, that in \cite{HRZ} the
anomalous equation for the left transforming currents (which can
be easily connected to ours) involving directly the physical
photon field is not properly implemented. Indeed, as we have shown
in Eq.~(\ref{anomaly}) at the fundamental quark level, we can
simply relate the axial anomaly between the physical photon
$\widetilde{A}$ and $A$.  This is due to the fact that we have an
Higgsing of the gauge symmetries.

The consequences of our result are two-fold. The first is that the
rate of the $\pi^0\rightarrow \widetilde{\gamma}
\widetilde{\gamma}$ is augmented by an order of magnitude
($N_c^2$) with respect to reference \cite{HRZ}. This fact can be
relevant for the physics of Quark-type stars featuring a CFL core.

The second is related to the solitonic (Skyrme) solutions of the
effective Lagrangian. In fact now we have the same winding number
as for ordinary QCD and hence we get massive excitations, which
after collective quantization, describe spin half particles with
the same quantum numbers of ordinary baryons. This is the actual
realization of the Quark-Hadron continuity \cite{SW} advocated for
the baryon-type sector of the Color-superconductive phase.

To the previous Lagrangian one can still add the extra Goldstone boson
associated to the $U_{V}(1)$ symmetry breaking without altering the previous
discussion (see \cite{CG}). One can check that the global anomalies are
correctly implemented by carefully gauging the Wess-Zumino term \cite
{Witten,anomalies} with respect to the flavor symmetries. Hence for the 3
flavor case too, the 't Hooft global anomalies are matched at finite (low
and high) density.

In writing the Goldstone Lagrangian we have not yet considered the
breaking of Lorentz invariance at finite density. Following
Ref.~\cite{CG} we note that the Goldstones obey, in medium, a
linear dispersion relation of the type $E={v}|{\vec{p}|}$, where
$E$ and $|{\vec{p}}|$ are respectively the energy and the momentum
of the Goldstone bosons. We can simply include the Lorentz
breaking by generalizing the effective Lagrangian (\ref{ef}) in
the following way:
\begin{equation}
L=\frac{F^{2}}{2}\,{\rm Tr}\left[\dot{U}\dot{U}^{\dagger} - v^2 {\vec{\nabla}
U \cdot \vec{\nabla}U^{\dagger}}\right]\ .
\end{equation}
Clearly by rescaling the vector coordinates $\vec{x}\rightarrow
\vec{x}/v$ we can recast the previous Lagrangian in the
Eq.~(\ref{ef}) form.

An important feature is that $\alpha$, being a differential form,
is unaffected by coordinate rescaling (actually topological terms
being independent on the metric are unaffected by medium effects),
and hence the Wess-Zumino term is not modified at finite matter
density.

Due to the breaking of the baryon number the final global symmetry
group in the superconductive phase differs from the ordinary
Goldstone phase.

\section{Electroweak for 3 flavors at high matter density}
\label{weak3}
 Next we
extend the previous effective Lagrangian by incorporating the
electroweak intermediate vector mesons as external fields. We
adopt a standard procedure which has been often employed in
literature. An example is the effective Lagrangian theories used
to describe a strong electroweak sector (technicolor like
theories) \cite{ARS}. A more closely related example is the
description of radiative and weak processes for low energy QCD at
zero temperature and matter density in the framework of Chiral
Perturbation Theory \cite{E}.

According to this procedure we need to partially gauge the flavor
subgroup by generalizing the effective Lagrangian in the following
way\footnote{ A word of caution is needed when describing non
leptonic weak decays. In this case we expect, as for the zero
density case, that the strong interactions may affect the
electroweak processes. Now one can first integrate out the heavy
electroweak intermediate vector meson fields by constructing a new
(weak) effective Lagrangian \cite{E}. The price to pay in this way
is the proliferation of unknown coefficients.}:
\begin{eqnarray}
DU &=&\partial U-i\frac{g}{\sqrt{2}}\left[ W^{+}\tau ^{+}+W^{-}\tau ^{-}
\right] U-i\frac{g}{\cos \theta _{W}}\,Z^{0}\,\left[ \tau ^{3}U-\sin
^{2}\theta _{W}\left[ Q,U\right] \right]  \nonumber \\
&&-ie\,A\,\left[ Q,U\right]  \nonumber \\
&=&\partial U-i\frac{g}{\sqrt{2}}\left[ W^{+}\tau ^{+}+W^{-}\tau ^{-}\right]
U-i\frac{g}{\cos \theta _{W}}\,Z^{0}\,\left[ \tau ^{3}U-\sin ^{2}\theta _{W}
\left[ Q,U\right] \right]  \nonumber \\
&&-i\widetilde{e}\,\widetilde{A}\,\left[ Q,U\right] -i\widetilde{e}\,\tan
\theta \widetilde{G}^{8}\,\left[ Q,U\right]  \label{elw}
\end{eqnarray}
where $\theta _{W}$ is the electroweak angle and
\begin{equation}
\tau ^{+}=\left(
\begin{array}{ccc}
0 & 1 & 0 \\
0 & 0 & 0 \\
0 & 0 & 0
\end{array}
\right) \ ,\qquad \tau ^{-}=\left(
\begin{array}{ccc}
0 & 0 & 0 \\
1 & 0 & 0 \\
0 & 0 & 0
\end{array}
\right) \ ,\qquad \tau ^{3}=\frac{1}{2}\left(
\begin{array}{ccc}
1 & 0 & 0 \\
0 & -1 & 0 \\
0 & 0 & -1
\end{array}
\right) \ .
\label{tau}
\end{equation}

The last two terms in Eq.~(\ref{elw}) describe the, non anomalous,
interaction of the Goldstone bosons with respectively the physical massless
photon and the physical massive gluon.

In the present framework the $SU_L(2)$ subgroup of the $SU_L(3)$
flavor group is completed gauged. At the quark level this allows
us to describe the weak interactions for the up and down quarks in
medium. Indeed it is easy to recognize that the upper
bidimensional sub-blocks of the matrices given in Eq.~(\ref{tau})
are the standard $SU_L(2)$ generators. {}For the strange quark we
considered just the diagonal interactions, conveniently encoded in
the last line of the matrix $\tau^3$. This is because the charm
quark is not included in the low energy Lagrangian we are
considering.

Clearly $\widetilde{e}=e\,\cos \theta $ is the finite density new
electric charge.

As for dilute QCD, when $U$ is evaluated on the vev we have a contribution
to the masses of the $W$ and $Z$ that we do not consider \cite{E}. We also
expect a non zero mixing among the gluons and the weak gauge bosons, which
will be explored in more detail for the 2 flavor case and that will be shown
to be small. One can study this mixing, in more detail, for the 3 flavor
case by generalizing the covariant derivative in Eq.~(\ref{cov1}) for the $
L/R$ fields.

For the massive quark case, we have to include the Cabibbo-Kobayashi-Maskawa
mixing angles. However it is straightforward to generalize the previous
expression to contain the mixing angles\footnote{
{}For example $\tau^{+}$ now becomes
\begin{equation}
\tau^{+}=\left(
\begin{array}{ccc}
0 & V_{ud} & V_{us} \\
0 & 0 & 0 \\
0 & 0 & 0
\end{array}
\right) \
\end{equation}
with $V$ mixing angles (see \cite{E}).}.

We finally observe that the electroweak physics for 3 flavors at high matter
density is similar to the zero density QCD except for a new photon electric
charge \cite{E} and the explicit presence of a massive gluon. The latter
couples to the Goldstone bosons via the standard charge operator $Q$.

\section{Review of the 2 flavor low energy Effective Theory}

\label{tre}

QCD with 2 flavors has gauge symmetry $SU_{c}(3)$ and global symmetry
\begin{equation}
SU_{L}(2)\times SU_{R}(2)\times U_{V}(1)\ .
\end{equation}
\noindent At very low baryon density it is reasonable to expect that the
confined Goldstone phase persists. However at very high density, it is seen,
via dynamical calculations \cite{ARW_98,RSSV_98}, that the ordinary
Goldstone phase is no longer favored compared with a superconductive one
associated with the following type of diquark condensates:
\begin{equation}
\langle L{^{\dagger }}^{a}\rangle \sim \langle \epsilon ^{abc}\epsilon
^{ij}q_{Lb,i}^{\alpha }q_{Lc,j;\alpha }\rangle \ ,\qquad \langle R{^{\dagger
}}^{a}\rangle \sim -\langle \epsilon ^{abc}\epsilon ^{ij}q_{Rb,i;\dot{\alpha}
}q_{Rc,j}^{\dot{\alpha}}\rangle \ ,
\end{equation}
$q_{Lc,i;\alpha }$, $q_{Rc,i}^{\dot{\alpha}}$ are respectively the
two component left and right spinors. $\alpha ,\dot{\alpha}=1,2$
are spin indices, $c=1,2,3$ stands for color while $i=1,2$
represents the flavor. If parity is not broken spontaneously, we
have
\begin{equation}
\left\langle L_{a}\right\rangle =\left\langle R_{a}\right\rangle
=f\delta _{a}^{3}\ ,  \label{vev}
\end{equation}
where we choose the condensate to be in the 3rd direction of color. The
order parameters are singlets under the $SU_{L}(2)\times SU_{R}(2)$ flavor
transformations while possessing baryon charge $\frac{2}{3}$. The vev leaves
invariant the following symmetry group:
\begin{equation}
\left[ SU_{c}(2)\right] \times SU_{L}(2)\times SU_{R}(2)\times \widetilde{U}_{V}(1)\ ,
\end{equation}
where $\left[ SU_{c}(2)\right] $ is the unbroken part of the gauge group.
The $\widetilde{U}_{V}(1)$ generator $\widetilde{B}$ is the following linear
combination of the previous $U_{V}(1)$ generator $B=\frac{1}{3}{\rm diag}(1,1,1)$ and the broken diagonal generator of the $SU_{c}(3)$ gauge group $T^{8}=\frac{1}{2\sqrt{3}}\,{\rm diag}(1,1,-2)$:\footnote{Nota Bene: for 2 flavors we keep the standard definition for the color
generators.}
\begin{equation}
\widetilde{B}=B-\frac{2\sqrt{3}}{3}T^{8}\ .  \label{residue}
\end{equation}
The quarks with color $1$ and $2$ are neutral under $\widetilde{B}$ and
consequently the condensate too\footnote{$\widetilde{B}$ is $\sqrt{2}\widetilde{S}$ of Ref.~\cite{CDS}}.

The superconductive phase for $N_{f}=2$ possesses the same global symmetry
group of the confined Wigner-Weyl phase \cite{S}.

The dynamics of the Godstone bosons can be efficiently encoded in a non
linear realization framework as presented in \cite{CDS}. In this framework
the relevant coset space is $G/H$ with $G=SU_{c}(3)\times U_{V}(1)$ and $H=SU_{c}(2)\times \widetilde{U}_{V}(1)$ is parameterized by \cite{CDS}:
\begin{equation}
{\cal V}=\exp (i\xi ^{i}X^{i})\ ,
\end{equation}
where $\{X^{i}\}$ $i=1,\cdots ,5$ belong to the coset space $G/H$ and are
taken to be $X^{i}=T^{i+3}$ for $i=1,\cdots ,4$ while
\begin{equation}
X^{5}=B+\frac{\sqrt{3}}{3}T^{8}={\rm diag}(\frac{1}{2},\frac{1}{2},0)\ .
\label{broken}
\end{equation}
$T^{a}$ are the standard generators of $SU(3)$. The coordinates
\begin{equation}
\xi ^{i}=\frac{\Pi ^{i}}{f}\quad i=1,2,3,4\ ,\qquad \xi
^{5}=\frac{\Pi ^{5}}{\widetilde{f}}\ ,
\end{equation}
via $\Pi $ describe the Goldstone bosons. The vevs $f$ and
$\widetilde{f}$ are expected, when considering asymptotically high
densities \cite{SW_98b}, to be proportional to $\mu $.

${\cal V}$ transforms non linearly:
\begin{equation}
{\cal V}(\xi )\rightarrow u_{V}\,g{\cal V}(\xi )h^{\dagger }(\xi ,g,u)h_{\widetilde{V}}^{\dagger }(\xi ,g,u)\ ,  \label{nl2}
\end{equation}
with
\begin{equation}
u_{V}\in U_{V}(1)\ ,\quad g\in SU_{c}(3)\ ,\quad h(\xi ,g,u)\in SU_{c}(2)\
,\quad h_{\widetilde{V}}(\xi ,g,u)\in \widetilde{U}_{V}(1)\ .
\end{equation}

The linear realizations are related via \cite{CDS} :
\begin{equation}
V_{a}=\frac{L_{a}+R_{a}}{\sqrt{2}}=\sqrt{2}\,f\,e^{i\frac{\Pi
^{5}}{\widetilde{f}}}\,{{\cal V}^{-1}}_{a}^{3}\ .
\end{equation}
The $V_{a}$ field explicitly describes the vev properties and, as expected,
transforms under the underlying gauge transformations as a diquark.

It is convenient to define the Hermitian (algebra valued) Maurer-Cartan
one-form
\begin{equation}
\omega _{\mu }=i{\cal V}^{\dagger }D_{\mu }{\cal V}\quad {\rm with}\quad
D_{\mu }{\cal V}=(\partial _{\mu }-ig_{s}G_{\mu }){\cal V}\ ,
\end{equation}
with gluon fields $G_{\mu }=G_{\mu }^{m}T^{m}$ while $g_{s}$ is the strong coupling constant. $\omega $ transforms as:
\begin{equation}
\omega _{\mu }\rightarrow h(\xi ,g,u)\omega _{\mu }h^{\dagger
}(\xi ,g,u)+i\,h(\xi ,g,u)\partial {_{\mu }}h^{\dagger }(\xi
,g,u)\ +i\,h_{\widetilde{V}}(\xi ,g,u)\partial _{\mu
}h_{\widetilde{V}}^{\dagger }(\xi ,g,u).
\end{equation}
Following \cite{CDS} we decompose $\omega _{\mu }$ into
\begin{equation}
\omega _{\mu }^{\parallel }=2S^{a}{\rm Tr}\left[ S^{a}\omega _{\mu }\right]
\quad {\rm and}\quad \omega _{\mu }^{\perp }=2X^{i}{\rm Tr}\left[
X^{i}\omega _{\mu }\right] \ ,
\end{equation}
where $S^{a}$ are the unbroken generators of $H$ with $S^{1,2,3}=T^{1,2,3}$,
$S^{4}=\widetilde{B}\,/\sqrt{2}$. Summation over repeated indices is assumed.

The most generic two derivative kinetic Lagrangian is \cite{CDS}.
\begin{equation}
L=f^{2}a_{1}{\rm Tr}\left[ \,\omega _{\mu }^{\perp }\omega ^{\mu
\perp }\, \right] +f^{2}a_{2}{\rm Tr}\left[ \,\omega _{\mu
}^{\perp }\,\right] {\rm Tr} \left[ \,\omega ^{\mu \perp
}\,\right] \ , \label{dt}
\end{equation}
The presence of a double trace term is due to the absence of the traceless
condition for the broken generator $X^{5}$ and it emerges naturally in the
non linear realization framework at the same order (in derivative expansion)
with respect to the single trace term. This is not the case for the linearly
realized effective Lagrangian \cite{R,ssh}.

\subsection{In Medium Fermions}

\label{fermions} Following Ref.~\cite{CDS} we define:
\begin{equation}
\widetilde{\psi}={\cal V}^{\dagger }\psi \ ,  \label{mq}
\end{equation}
transforming as $\widetilde{\psi}\rightarrow h_{\widetilde{V}}(\xi,g,u)h(\xi
,g,u)\widetilde{ \psi}$ and $\psi$ possesses an ordinary quark
transformations (as Dirac spinor). This construction mimics the Heavy Quark
Effective formalism \cite{HQ}.

The whole \cite{CDS} non linearly realized effective Lagrangian describing
in medium fermions, gluons and their self interactions, up to two
derivatives is:
\begin{eqnarray}
{\cal L}=~ &&f^{2}a_{1}{\rm Tr}\left[ \,\omega _{\mu }^{\perp
}\omega ^{\mu \perp }\,\right] +f^{2}a_{2}{\rm Tr}\left[ \,\omega
_{\mu }^{\perp }\,\right] {\rm Tr}\left[ \,\omega ^{\mu \perp
}\,\right]  \nonumber \\ &+&b_{1}\overline{\widetilde{\psi
}}i\gamma ^{\mu }(\partial _{\mu }-i\omega _{\mu }^{\parallel
})\widetilde{\psi }+b_{2}\overline{\widetilde{\psi }} \gamma ^{\mu
}\omega _{\mu }^{\perp }\widetilde{\psi }  \nonumber \\
&+&m_{M}\overline{\widetilde{\psi }^{C}}_{i}\gamma
^{5}(iT^{2})\widetilde{\psi }_{j}\varepsilon ^{ij}+{\rm h.c.}\ ,
\end{eqnarray}
where $\widetilde{\psi }^{C}=i\gamma ^{2}\widetilde{\psi }^{\ast }$, $i,j=1,2 $ are flavor indices and
\begin{equation}
T^{2}=S^{2}=\frac{1}{2}\left(
\begin{array}{ll}
\sigma ^{2} & 0 \\
0 & 0
\end{array}
\right) \ ,
\end{equation}
$a_{1},~a_{2},~b_{1}$ and $b_{2}$ are real coefficients while
$m_{M}$ is complex.
 {}For later convenience, it is convenient, to
express the third and forth terms as
\begin{equation}
b_{1}\overline{\widetilde{\psi }}i\gamma ^{\mu }(\partial _{\mu }-i\omega
_{\mu })\widetilde{\psi }+(b_{2}-b_{1})\overline{\widetilde{\psi }}\gamma
^{\mu }\omega _{\mu }^{\perp }\widetilde{\psi }=b_{1}\overline{{\psi }}
i\gamma ^{\mu }{D}_{\mu }{\psi }+\left( b_{2}-b_{1}\right) \overline{\widetilde{\psi }}\gamma ^{\mu }\omega _{\mu }^{\perp }\widetilde{\psi }.
\end{equation}

{}From the last two terms, representing a Majorana mass term for the quarks,
we deduce that the massless degrees of freedom are the $\psi _{a=3,i}$ which
possess the correct quantum numbers to match the 't~Hooft anomaly conditions
\cite{S}. To the previous general effective Lagrangian we should also add
the $SU(2)$ gluon kinetic term.

Following Ref.~\cite{CG} the breaking of Lorentz invariance to the $O(3)$
subgroup can easily be taken into account by providing different
coefficients to the temporal and spatial indices of the Lagrangian, i.e.:
\begin{eqnarray}
{\cal L}=~ &&f^{2}a_{1}{\rm Tr}\left[ \,\omega _{0}^{\perp }\omega
_{0}^{\perp }-{\alpha }_{1}\vec{\omega}^{\perp
}\vec{\omega}^{\perp }\, \right] +f^{2}a_{2}\left[ {\rm Tr}\left[
\,\omega _{0}^{\perp }\,\right] {\rm Tr}\left[ \,\omega
_{0}^{\perp }\,\right] -{\alpha }_{2}{\rm Tr}\left[
\,\vec{\omega}^{\perp }\,\right] {\rm Tr}\left[
\,\vec{\omega}^{\perp }\, \right] \right]  \nonumber \\
&+&b_{1}\overline{\widetilde{\psi }}i\left[ \gamma ^{0}(\partial
_{0}-i\omega _{0}^{\parallel })+\beta _{1}\vec{\gamma}\cdot \left(
\vec{ \nabla}-i\vec{\omega}^{\parallel }\right) \right]
\widetilde{\psi }+b_{2} \overline{\widetilde{\psi }}\left[ \gamma
^{0}\omega _{0}^{\perp }+\beta _{2} \vec{\gamma}\cdot
\vec{\omega}^{\perp }\right] \widetilde{\psi } \nonumber \\
&+&m_{M}\overline{\widetilde{\psi }^{C}}\gamma
^{5}(iT^{2})\widetilde{\psi }+ {\rm h.c.}\ ,
\end{eqnarray}
where the new coefficients $\alpha $s and $\beta $s encode the effective
breaking of Lorentz invariance and the flavor indices are omitted.

\section{Electroweak interactions in matter}
\label{quattro}

To construct the low energy effective theory for the electroweak sector we
need to gauge the weak isospin as well as the hypercharge sector and finally
identify the correct massless as well massive eigenstates.

We identify the $SU_{L}(2)$ generators $\tau _{L}^{i}=\sigma _{L}^{i}/2$
with $i=1,2,3$ and $\sigma ^{i}$ the standard Pauli's matrices as the weak
generators. The hypercharge is, in full generality, $\displaystyle{Y=\tau
_{R}^{3}+ \frac{B-L}{2}}$ with $L$ the lepton number while $\tau _{R}$
labels the $SU_{R}(2)$ generators. Quarks and leptons have standard charges,
in particular $B=1/3$ for quarks.

We need to generalize the one form $\omega_{\mu}=i{\cal V}^{\dagger}D_{\mu}
{\cal V}$ by introducing the new covariant derivative:
\begin{equation}
D_{\mu }{\cal V}=(\partial _{\mu }-ig_s G_{\mu }-ig^{\prime} Y\, B^y_{\mu})
{\cal V}=(\partial _{\mu }-ig_s G_{\mu }-ig^{\prime} \frac{B}{2}\,
B^y_{\mu}) {\cal V} \ .  \label{newD}
\end{equation}
$B^y_{\mu}$ is the standard hypercharge gauge field. Neglecting, for the
moment, the breaking of Lorentz invariance and substituting Eq.~(\ref{newD})
in Eq.~(\ref{dt}) we have the following quadratic terms:
\begin{equation}
a_1 g_s^2 \frac{v^2}{2} \sum_{i=4}^7 G_{\mu}^{i} G^{i \mu} +
\left( a_1 + 2a_2\right) \frac{f^2}{2} \left[\frac{g_s}{\sqrt{3}}
G^8_{\mu} + \frac{ g^{\prime}}{3} B^y_{\mu} \right]^2 \ .
\label{quad}
\end{equation}

We now rewrite the previous terms using the electroweak eigenstates
associated with the photon field $A_{\mu }$ and the neutral massive vector
boson $Z_{\mu }^{0}$.
\begin{equation}
B_{\mu }^{y}=\cos \theta _{W}A_{\mu }-\sin \theta _{W}Z_{\mu }^{0}\ ,
\end{equation}
with $\theta _{W}$ the standard electroweak angle. Focusing only on the
second term in Eq.~(\ref{quad}) one has:
\begin{eqnarray}
\left( a_{1}+2a_{2}\right) \frac{f^{2}}{2} &~&\left\{ \left[
\frac{g_{s}}{\sqrt{3}}G^{8}+\frac{e}{3}A\right] ^{2}\right.
\nonumber \\ &~&\left. -\frac{2}{3}\,e\,\tan \theta
_{W}Z^{0}\left[ \frac{g_{s}}{\sqrt{3}} G^{8}+\frac{e}{3}A\right]
+\tan ^{2}\theta _{W}\frac{{e}^{2}}{9}{Z^{0}} ^{2}\right\} \ ,
\end{eqnarray}
where $e=g^{\prime }\cos \theta _{W}$ is the standard electric charge and
for simplicity we have dropped the Lorentz indices. To this term we have to
add the electroweak quadratic mass term for $Z^{0}$
\begin{equation}
{\frac{1}{2}}m_{Z}^{2}{Z^{0}}^{2}.
\end{equation}

The new massless eigenstate is interpreted as the, in medium, photon and is
given by:
\begin{equation}
\widetilde{A}_{\mu }=\cos \theta _{Q}A_{\mu }-\sin \theta _{Q}G_{\mu }^{8}\ ,
\end{equation}
with
\begin{equation}
\cos \theta _{Q}=\sqrt{3}\frac{g_{s}}{\sqrt{3g_{s}^{2}+e^{2}}}\ ,\qquad \sin
\theta _{Q}=\frac{e}{\sqrt{3g_{s}^{2}+e^{2}}}\ .
\end{equation}
The massive state orthogonal to $\widetilde{A}_{\mu}$ is:
\begin{equation}
\widetilde{G}_{\mu }^{8}=\cos \theta _{Q}G_{\mu }^{8}+\sin \theta _{Q}A_{\mu
}\ .
\end{equation}
Using the new base, the complete tree level quadratic term involving $\widetilde{G}^{8}$ and $Z^{0}$ is
\begin{eqnarray}
\left( a_{1}+2a_{2}\right) \frac{f^{2}}{18} &~&\left\{ \left(
3g_{s}^{2}+e^{2}\right) \widetilde{G}^{82}-{2}\,e\,\tan \theta
_{W}Z^{0} \widetilde{G}^{8}\sqrt{3g_{s}^{2}+e^{2}}+\tan ^{2}\theta
_{W}\,{e^{2}}\,{\ Z^{0}}^{2}\right\}  \nonumber \\
&+&\frac{m_{Z}^{2}}{2}{Z^{0}}^{2}\ .
\end{eqnarray}
By diagonalizing the previous matrix we have the new massive eigenstates:
\begin{eqnarray}
\hat{G}_{\mu }^{8} &=&\cos \theta _{M}\widetilde{G}_{\mu }^{8}+\sin \theta
_{M}Z_{\mu }^{0}\ ,  \nonumber \\
\widetilde{Z}_{\mu }^{0} &=&\cos \theta _{M}Z_{\mu }^{0}-\sin \theta _{M}
\widetilde{G}_{\mu }^{8},
\end{eqnarray}
with
\begin{equation}
\tan 2\theta _{M}=\frac{2f^{2}\,\tan \theta _{W}\,\left(
a_{1}+2a_{2}\right) \sqrt{3g_{s}^{2}+e^{2}}}{\left(
a_{1}+2a_{2}\right) f^{2}\left[ \tan ^{2}\theta _{W}e^{2}-\left(
3g_{s}^{2}+e^{2}\right) \right] +9m_{Z}^{2}} \approx
\frac{2f^{2}\,\tan \theta _{W}}{9m_{Z}^{2}}\left(
a_{1}+2a_{2}\right) \sqrt{3g_{s}^{2}+e^{2}}\ ,
\end{equation}
where in the last step we have considered the physical limit
$m_{Z}^{2}\gg f^{2}$. In the same limit, as expected,
$\widetilde{G}^{8}$ and $Z^{0}$ do not mix much and we can use
them as physical eigenstates.

Having identified the correct physical eigenvalues and eigenvectors we are
now ready to consider the Lagrangian for the quarks. The relevant Lagrangian
terms are:
\begin{equation}
b_{1}\overline{{\psi }}i\gamma ^{\mu }{D}_{\mu }{\psi }+\left(
b_{2}-b_{1}\right) \overline{\widetilde{\psi }}\gamma ^{\mu
}\omega _{\mu }^{\perp }\widetilde{\psi }\ , \label{kinetic2}
\end{equation}
where we generalize the covariant derivative to describe the weak
interactions in the following way:
\begin{eqnarray}
{D}_{\mu } &=&\partial _{\mu }-i\frac{g}{\sqrt{2}}\left[ W_{\mu }^{+}\tau
_{L}^{+}\times {\mbox{\bf 1}}+W_{\mu }^{-}\tau _{L}^{-}\times {\mbox{\bf 1}}
\right] -i\frac{g}{\cos \theta _{W}}Z_{\mu }^{0}\left[ \tau _{L}^{3}\times {
\mbox{\bf 1}}-\sin \theta _{W}^{2}Q\times {\mbox{\bf 1}}\right]  \nonumber \\
&&-ieA_{\mu }Q\times {\mbox{\bf 1}}-ig_{s}\sum_{m=1}^{8}G_{\mu }^{m}{\mbox{\bf 1}\times }T^{m}\ .  \label{standard}
\end{eqnarray}
We adopted a notation similar to the one used in the $3$ flavor case; i.e.
flavor$_{2\times 2}\times $ color$_{3\times 3}$. We also have
\begin{equation}
Q={\tau ^{3}+\frac{B-L}{2}}\ ,
\end{equation}
with $\tau =\tau _{L}+\tau _{R}$. Considering the gluon-photon mixing we
have:
\begin{equation}
D_{\mu }=\cdots -i\widetilde{e}\widetilde{A}_{\mu }\widetilde{Q}-i\frac{g_{s}}{\cos \theta _{Q}}\widetilde{G}_{\mu }^{8}\left[ \cos \theta _{Q}^{2}{\mbox{\bf 1}\times }T^{8}+\sqrt{3}\sin \theta _{Q}^{2}Q\times {\mbox{\bf 1}}
\right] \ ,
\end{equation}
where the dots represent the terms unchanged in Eq.~(\ref{standard}). $\tilde{e}$ is the new electric charge
\begin{equation}
\widetilde{e}=e\,\cos \theta _{Q}\ ,
\end{equation}
$\widetilde{Q}$ is the new electric charge operator associated with the
field $\widetilde{A}_{\mu }$:
\begin{equation}
\widetilde{Q}={\tau ^{3}\times {\mbox{\bf 1}}+\frac{\tilde{B}-L}{2}}=Q\times
{\mbox{\bf
1}}-\frac{1}{\sqrt{3}}{\mbox{\bf 1}\times }T^{8}\ ,  \label{newcharge}
\end{equation}
The quarks that acquires a mass term (i.e. the ones in the color direction one and two) have half integer charges under $\widetilde{Q}$  while
the massless quarks (the ones in direction three of color) have the ordinary
proton and neutron charges in units of $\widetilde{e}$.

To the leading order the term containing $\omega ^{\perp }$, in
Eq.~(\ref{kinetic2}), does not involve any electric interaction
but it does affect the neutral weak currents. Expanding around
${\cal V}=1$ one gets:
\begin{equation}
(b_{2}-b_{1})\overline{\psi }\gamma ^{\mu }\left( g_{s}\sum_{m=4}^{7}G_{\mu
}^{m}T^{m}+\frac{g_{s}}{\sqrt{3}\cos \theta _{Q}}\widetilde{G}_{\mu
}^{8}X^{5}-\frac{g}{3}\frac{\sin \theta _{W}^{2}}{\cos \theta _{W}}Z_{\mu
}^{0}X^{5}\right) \psi +\cdots \ ,
\end{equation}
where the dots stand for terms involving higher terms in the
expansion of ${\cal V}$. The neutral weak current is directly
affected by finite density effects even when neglecting the small
physical mixing between the eighth gluon $\widetilde{G}^{8}$ and
$Z^{0}$. We also observe that the modified electroweak coupling
only emerges for quarks with color indices 1 and 2, since
$\displaystyle{X^{5}=\frac{1}{2}{\rm diag}(1,1,0)}$. The new term
does not affect the light quarks with color index 3. In particular
this effect depends crucially upon the unknown ratio:${\left(
b_{2}-b_{1}\right) } /{b_{1}}$, and depends on $\mu $. The full
modified quark coupling to the neutral weak current is now:
\begin{equation}
b_{1}\frac{g}{\cos \theta _{W}}\,Z_{\mu }^{0}\bar{\psi}\gamma
^{\mu }\left[ T_{L}^{3}-\sin \theta
_{W}^{2}Q-\frac{b_{2}-b_{1}}{3\,b_{1}}\sin \theta
_{W}^{2}(B+Q-\widetilde{Q})\right] \psi \ , \label{z0}
\end{equation}
where we used Eq.~(\ref{newcharge}) and $X^{5}=B+Q-\widetilde{Q}$.

The modification of the neutral current coupling to the quarks at
high matter density constitutes our main result for what concerns
the 2 flavors case.

We can point immediately to relevant phenomenological consequences
of our result. The cooling history of compact objects, like for
instance the ones generated as remnant (proto-star) of a Supernova
explosion \cite{Shapiro}, is heavily related to the neutrino
physics. The neutrino transparency of a given star \cite{Shapiro}
can significantly modify the cooling history and is associated
with the mean free path of a neutrino in the star. Indeed for
ordinary neutron stars \cite{Shapiro} the mean free path is
strongly affected by the $neutron-{\nu}$ scattering reaction due
to the presence of neutral currents. In a 2SC star we then expect
the $quark-{\nu}$ scattering to play an equally important role and
its modification with respect to the zero density case to be
dictated by the new weak coupling of Eq.~(\ref{z0}).

Finally, on general grounds, we have shown that a direct
phenomenological evidence for color superconductivity can be, in
principle, found by studying the neutral weak interactions
associated with a given high matter density physical system.

\section{Conclusions}

\label{cinque}

We investigated the electroweak interactions for the relevant
color superconductivity phases. More specifically we included the
$QED$ interactions at the effective Lagrangian level for the three
flavor case at high density developed in \cite{CG}. We identified
the physical photon and provided the Wess-Zumino term needed for
describing the $\pi ^{0}\rightarrow \gamma \gamma $ interaction in
medium.

We found, as main result for the 3 flavor case, that at high
matter density the quasiparticle Goldstone boson $\pi^0$ decay
into two physical massless photons is identical to the zero
density case once we replace the electric charge with the modified
charge $\widetilde{e}=e\, \cos\theta$ and also use the new decay
constant.  This result is consistent with the underlying global
anomaly constraints at finite density \cite{S,HSaS}. However it
differs from the conclusions drawn in \cite{HRZ}. We finally
extended the effective Lagrangian to include the weak
interactions.

{}For the 2 flavor case we constructed the effective Lagrangian
theory containing the proper electroweak theory. We, first,
identified the physical photon. We have also discovered that our
Lagrangian not only correctly yields the physical eigenstates for
the photon and the eight gluon but also predicts that $Z^{0}$ has
a modified  coupling to the quarks (in the direction 1 and 2 of
color) with respect to the zero density case. The physical effect
is not suppressed with respect to zero density weak processes. The
effect, which is our main result for the 2 flavor case, is not
suppressed with respect to the zero density weak processes. On
general grounds we expect this result to affect the Quark-stars
(with a 2SC component) cooling processes \cite{Shapiro} via
neutrino emission. The neutrino transparency question is indeed
directly related to the mean free path of a neutrino in a compact
star. {}In neutron stars \cite{Shapiro} the mean free path is
known to be significantly affected by the $neutron-{\nu}$
scattering reaction due to neutral currents. Likewise in a 2SC
star the $quark-{\nu}$ scattering will play a relevant role with
the new weak coupling displayed in Eq.~(\ref{z0}). Clearly a
dynamical computation of the quantity $\left( b_{2}-b_{1}\right)
/b_{1}$, although beyond the goal of this paper, might be very
interesting.

\vskip2cm \centerline{\bf Acknowledgments}

It is a pleasure for us to thank J. Schechter for interesting
discussions and careful reading of the manuscript. We also thank
P. Hoyer, J. Lenaghan and R. Ouyed for helpful discussions and
careful reading of the manuscript. The work of Z.D. is partially
supported by the US DOE under contract DE-FG-02-92ER-40704.


\end{document}